\documentclass[12pt]{article}
\usepackage[cp1251]{inputenc}
\usepackage{amsmath}
\usepackage{amsfonts}
\usepackage{amssymb}
\def\pa{\partial}                       
\def\beq{\begin{eqnarray}}    
\def\eeq{\end{eqnarray}}      

\begin{document}
\date{}
\begin{center}
{\Large\textbf{Gauge-invariant models of interacting fields with spins 3,1 and 0
}}

\vspace{18mm}

{\large
P.M. Lavrov$^{(a, b)} \footnote{E-mail: lavrov@tspu.edu.ru}$
}

\vspace{8mm}

\noindent  ${{}^{(a)}} ${\em
Center of Theoretical Physics, \\
Tomsk State Pedagogical University,\\
Kievskaya St.\ 60, 634061 Tomsk, Russia}

\noindent  ${{}^{(b)}} ${\em
National Research Tomsk State  University,\\
Lenin Av.\ 36, 634050 Tomsk, Russia}

\vspace{20mm}

\begin{abstract}
\noindent
New local gauge-invariant models of interacting fields with spins 3, 1 and 0 are found.
The construction of the models is completely based on the new approach to the deformation
problem proposed in our papers (Buchbinder and Lavrov in JHEP 06: 097, 2021;
Buchbinder and Lavrov in  Eur. Phys. J. C 81:856, 2021;
Lavrov in Eur. Phys. J. C 82:429, 2022). The approach allows to describe the deformation
procedure for theories with gauge freedom in an explicit and closed form in terms of
special anticanonical transformations of the Batalin-Vilkovisky formalism.
The action of the models appears as a local part of the deformed action which is represented
by a functional of the fourth order in fields. This functional is invariant under
original gauge transformations.

\end{abstract}

\end{center}

\vfill

\noindent {\sl Keywords:} BV-formalism, deformation procedure,
anticanonical transformations
\\

\noindent PACS numbers: 11.10.Ef, 11.15.Bt
\newpage

\section{Introduction}
It is well-known  that the Batalin-Vilkovisky (BV) formalism \cite{BV,BV1} is the powerful
method not only for quantization of general gauge theories but an important tool
for solving different problems on classical and quantum levels.
So, almost thirty years ago it was proposed to include the deformation procedure
which is considered  as a way to construct suitable interactions
between fields, in solutions to the classical master-equation of the BV formalism \cite{BH,H}.
It was proposed to search solutions in the form of Taylor expansion with respect
to a deformation parameter. In general, it causes infinite series  of equations which
are analysed with the help of cohomological methods
(for recent application and references see, for example, \cite{FKSS}).
Recently,
a new approach to find solutions of the deformation procedure within the BV-formalism
has been proposed \cite{BL-1,BL-2,L-2022}. The approach uses the fact that
the classical master-equation  is formulated in terms of the antibracket. In turn,
the antibracket obeys the property of invariance under anticanonical transformations.
In particular, it means that any two solutions of the classical master-equation can be
connected to each other by an appropriate anticanonical transformation.
In \cite{BL-1,BL-2,L-2022}, it was proved that the deformation procedure can be solved
with the help of special anticanonical transformations acting non-trivially in the sector of
initial fields of a given gauge theory. In turn, it allows to describe
the deformation procedure in an explicit and closed form.
The deformation looks like the result of an explicit summation of a series of the
standard approach based on cohomological methods. In fact, it opens new possibilities
in construction of gauge-invariant models of interacting fields.
Using new approach,  in \cite{L2}, the first new gauge-invariant model
with interaction has been constructed. This model presents the massless spin 3 field
interacting with a real scalar field. The action of the model is a local functional of the
fourth order in fields which is invariant under original gauge transformations.
This model is of particular interest due to the fact that it cannot be reproduced
using the standard Noether procedure, which
is accepted as the main method for describing interactions
in the theory of higher spin fields \cite{BLS,FIPT,Zinov,KMP}. In connection with the locality
problem\footnote{For quite
recent discussions of the locality problem in the theory of higher spin fields
see \cite{Vas,Did} and references therein.}
for interactions
of higher spin field, the model may have some interest because of the local nature of quartic
vertices.

In this paper, we continue studying the interaction of a massless spin 3 field
with a real scalar field for the case when a massive vector field is additionally involved.
As in \cite{L2}, the interaction appears due to deformations of initial gauge action under
special anticanonical transformations \cite{BL-1,BL-2,L-2022}.
 The initial action is sum of the
Fronsdal action for massless spin 3 fields \cite{Fronsdal-1} and the actions for
a massive vector field and a  real scalar field. The generating function of anticanonical
transformations is chosen in a form that preserves gauge symmetry after deformations.
In turn, it allows to extract a local functional from the deformed action  which is
invariant under original gauge transformations. This functional defines explicitly
the gauge-invariant model and depends on fields up to the fourth order. By passing,
it is proved that cubic vertices invariant under
original gauge transformations are forbidden.

The plan of the paper is as follows. Section 2 contains a formulation of an initial
free gauge model of massless spin 3 field and the massive vector and scalar fields.
It is shown that local cubic vertices invariant under original gauge transformations
are forbidden. In section 3, local gauge-invariant quartic vertices are constructed.
In section 4, a new local gauge-invariant model with mixed quartic vertices is proposed.
Section 5 is summarized the results.

The DeWitt's condensed notations are systematically used \cite{DeWitt}.
The right functional derivatives are marked by special symbols $"\leftarrow"$.
Arguments of any functional
are enclosed in square brackets $[\;]$, and arguments of any
function are enclosed in parentheses, $(\;)$.

\section{Cubic vertices}
Our starting point is the free model of massless spin $3$,
vector  and scalar fields in flat Minkowski space of dimension $d$ with the action
\beq
\label{c1}
S_0[\varphi,A,\phi]=S_0[\varphi]+S_0[A]+S_0[\phi],
\eeq
where $S_0[\varphi]$ is the Fronsdal action for completely symmetric third rank tensor
$\varphi^{\mu\nu\lambda}=\varphi^{\mu\nu\lambda}(x)$,
\beq
\nonumber
S_0[\varphi]&=&\int dx \Big(\varphi^{\mu\nu\rho}\Box\varphi_{\mu\nu\rho}-
3\eta_{\mu\nu}\eta^{\rho\sigma}\varphi^{\mu\nu\delta}
\Box\varphi_{\rho\sigma\delta}-
\frac{3}{2}\varphi_{\mu\nu\lambda}\eta^{\mu\nu}\pa^{\lambda}\pa_{\alpha}
\varphi^{\alpha\beta\gamma}\eta_{\beta\gamma}-\\
\label{b1}
&&\qquad\qquad
-3\varphi^{\mu\rho\sigma}\pa_\mu\pa^\nu\varphi_{\nu\rho\sigma}+
6\eta_{\mu\nu}\varphi^{\mu\nu\delta}\pa^\rho\pa^\sigma
\varphi_{\rho\sigma\delta}\Big),
\eeq
$S_0[A]$ and  $S_0[\phi]$ are the actions of
a massive vector field $A^{\mu}=A^{\mu}(x)$ and
  real scalar field $\phi=\phi(x)$, respectively,
\beq
\label{c1a}
S_0[A]=\int dx \Big(-\frac{1}{4}F_{\mu\nu}F^{\mu\nu} -\frac{1}{2}m_0^2A_{\mu}A^{\mu}
\Big),\quad
S_0[\phi]=\frac{1}{2}\int dx \big( \pa_{\mu}\phi\;\pa^{\mu}\phi-m^2\phi^2\big) .
\eeq
The action (\ref{c1}) is invariant under the following gauge transformations
\beq
\label{c2}
\delta\varphi^{\mu\nu\lambda}=\pa^{(\mu}\xi^{\nu\lambda)},\quad
\delta A^{\mu}=0,\quad
\delta\phi=0, \quad
\eta_{\mu\nu}\xi^{\mu\nu}=0,
\eeq
where gauge parameters $\xi^{\nu\lambda}$ are arbitrary symmetric functions of space-time
coordinates.
As it was explained in \cite{L2}, algebra of such gauge transformations is Abelian and
belongs
to the class of first-stage reducible theories in the terminology of BV-formalism.

Now, we consider the special class of  deformations of initial action using
the approach proposed in our papers \cite{BL-1,BL-2,L-2022} when anticanonical transformations
of the BV-formalism are ruled out the gauge-invariant deformations of an
initial gauge theory.
 Here, we restrict ourself to the case of
anticanonical transformations acting effectively in the sector of fields
$\varphi^{\mu\nu\lambda}$ only. It means that the generating function of anticanonical
transformations should be  completely symmetric third rank tensor
$h^{\mu\nu\lambda}=h^{\mu\nu\lambda}(\varphi,A,\phi)$.
 In construction of suitable generating
functions
$h^{\mu\nu\lambda}$, we have to take
into account the dimensions of quantities involved
in the initial action $S_0[\varphi,A,\phi]$,
\beq
\label{c3}
{\rm dim}(\varphi^{\mu\nu\lambda})={\rm dim}(A^{\mu})={\rm dim}(\phi)=\frac{d-2}{2},\;
{\rm dim}(\xi^{\mu\nu})=\frac{d-4}{2},\; {\rm dim}(\pa_{\mu})=1, \;
{\rm dim}(\Box)=2.
\eeq
The generating function $h^{\mu\nu\lambda}$ should be  non-local  with
the dimension equals to $(d-2)/2$. The non-locality is achieved by using the operator
$1/\Box$. Here, we shall be interested in cubic vertices of the form $\sim \varphi A\phi$
 to continue our knowledge about cubic vertices in theories of spin 3 fields \cite{L2}.
It means that the $h^{\mu\nu\lambda}$ can be constructed with the help
of terms proportional
to $\sim A\phi$.
 The tensor structure of $h^{\mu\nu\lambda}$
is obeyed by using
partial derivatives $\pa_{\mu}$, the metric tensor
$\eta_{\mu\nu}$ and vector fields $A^{\mu}$.
The minimal number of derivatives equals to 2. Therefore,
the more general form of
$h^{\mu\nu\lambda}=h^{\mu\nu\lambda}(A,\phi)$ satisfying requirements listed above reads
\beq
\nonumber
&&h^{\mu\nu\lambda}=a_0\frac{1}{\Box}
\big(c_1\pa^{(\mu}\pa^{\nu}A^{\lambda)} \;\phi+
c_2\pa^{(\mu}\pa^{\nu}\phi\;A^{\lambda)}+
c_3\pa^{(\mu}\phi\;\pa^{\nu}A^{\lambda)}+
c_4\eta^{(\mu\nu}\Box A^{\lambda)}\;\phi
+\\
\label{c4}
&&\qquad\qquad\quad+c_5\Box\phi\;\eta^{(\mu\nu}A^{\lambda)}+
c_6\eta^{(\mu\nu}\pa_{\sigma}A^{\lambda)}\;\pa^{\sigma}\phi+
c_7\eta^{(\mu\nu}\pa^{\lambda)}\phi\;\pa^{\sigma}A_{\sigma}\big),
\eeq
where $a_0$ is the coupling constant with ${\rm dim}(a_0)=-(d-2)/2$
and $c_i,\;\; i=1.2,...,7$ are arbitrary dimensionless constants.
Local part of the deformed action has the form
\beq \label{c5}
S_{loc}[\varphi,A,\phi]=S_0[\varphi,A,\phi]+S_{1\;loc}[\varphi,A,\phi]
\eeq
where
$S_{1\; loc}=S_{1\;
loc}[\varphi,A,\phi]$,
\beq
\nonumber
&&S_{1\; loc}=2a_0\int dx \varphi_{\mu\nu\lambda}\Big[c_1\pa^{(\mu}\pa^{\nu}A^{\lambda)}
\;\phi+ c_2\pa^{(\mu}\pa^{\nu}\phi\;A^{\lambda)}+
c_3\pa^{(\mu}\phi\;\pa^{\nu}A^{\lambda )}-\\
\nonumber
&&\qquad\qquad\qquad-
\big(c_1+c_4(d+1)\big)\eta^{(\mu\nu}\Box A^{\lambda)}\;\phi
-\big(c_2+c_5(d+1)\big)\Box\phi\;\eta^{(\mu\nu}A^{\lambda)}-\\
\nonumber
&&\qquad\qquad\qquad-
\big(c_3+c_6(d+1)\big)\eta^{(\mu\nu}\pa_{\sigma}A^{\lambda)}\;\pa^{\sigma}\phi -
\big(c_3+c_7(d+1)\big)\eta^{(\mu\nu}\pa^{\lambda)}\phi\;\pa_{\sigma}A^{\sigma}-\\
&&\qquad\qquad\qquad-2c_1\eta^{(\mu\nu}\pa^{\lambda)}\pa^{\sigma}A_{\sigma}\;\phi-
2c_2\eta^{(\mu\nu}\pa^{\lambda)}\pa^{\sigma}\phi\;A_{\sigma}\Big],
\label{c6}
\eeq
corresponds to possible cubic vertices. Due to the special structure of generating functions
of anticanonical transformations, the matrix $(M^{-1})^i_{\;j}$ (for details see
\cite{BL-1,L-2022})
can be found in the explicit form
\beq
\label{M}
(M^{-1})^i_{\;j}=\left(\begin{array}{ccc}
                     E^{\mu\nu\lambda}_{\rho\sigma\gamma}
                     &-h^{\mu\nu\lambda}(A,\phi)\overleftarrow{\pa}_{\!A^{\alpha}}&
                     -h^{\mu\nu\lambda}(A,\phi)\overleftarrow{\pa}_{\!\phi}\\
                     0& \delta^{\mu}_{\alpha}& 0\\
                     0&0&1
                     \end{array}\right),
\eeq
where $E^{\mu\nu\lambda}_{\rho\sigma\gamma}$ are elements of the unit matrix in the space of
third rank symmetric tensors.
From the structure of matrix (\ref{M}), it follows that  the deformed gauge generators
coincide
with initial ones.

Consider the variation of $S_{1\; loc}$
under the gauge transformations (\ref{c2}). We obtain the result
\beq
\nonumber
&&\delta S_{1\; loc}=-6a_0\int dx \xi_{\nu\lambda}\Big[-
2c_4(d+1)\Box \pa^{\nu}A^{\lambda}\;\phi
+(2c_1-c_3-2c_6(d+1))\pa^{\sigma}\pa^{\nu}A^{\lambda}\;\pa_{\sigma}\phi-\\
\nonumber
&&\qquad\qquad\qquad\quad-3c_1\pa_{\sigma}\pa^{\nu}\pa^{\lambda}A^{\sigma}\;\phi +
c_1\pa^{\nu}\pa^{\lambda}A^{\sigma}\;\pa_{\sigma}\phi -
2c_5(d+1)\Box\pa^{\nu}\phi\;A^{\lambda}+\\
\nonumber
&&\qquad\qquad\qquad\quad+
(2c_2-c_3-2c_6(d+1))\pa^{\lambda}\pa^{\sigma}\phi\;\pa_{\sigma}A^{\nu}-
3c_2\pa^{\nu}\pa^{\lambda}\pa^{\sigma}\phi\;A_{\sigma}+\\
\nonumber
&&
\qquad\qquad\qquad\quad+(c_2-2c_3-2c_7(d+1))
\pa^{\nu}\pa^{\lambda}\phi\;\pa_{\sigma}A^{\sigma}+(c_3-2c_2-2c_5(d+1))
\Box\phi\;\pa^{\nu}A^{\lambda}+\\
\nonumber
&&\qquad\qquad\qquad\quad+(c_3-2c_1-2c_4(d+1))\Box A^{\nu}\;\pa^{\lambda}\phi+
(c_3-4c_2)\pa_{\sigma}\pa^{\nu}\phi\;\pa^{\lambda}A^{\sigma}-\\
&&\qquad\qquad\qquad\quad-(c_3+4c_1+2c_7(d+1))
\pa^{\nu}\phi\;\pa^{\lambda}\pa_{\sigma}A^{\sigma}\Big].
\label{c13}
\eeq
The requirement $\delta S_{1 loc}=0$ leads to the conditions
\beq
\label{c15}
c_i=0,\;\; i=1,2,...,7.
\eeq
It means that
\beq
\label{c16}
S_{1\;loc}[\varphi,A,\phi]=0.
\eeq
Therefore, we have proved impossibility in the theory of massless  spin 3 field
interacting with
massive vector and scalar fields to construct  cubic
vertices $\sim\varphi A\phi$ being invariant under initial gauge transformations (\ref{c2}).
In this point, the situation is similar with cubic vertices studied in \cite{L2}.

\section{Quartic vertices}
In the paper \cite{L2},  it was proved that although gauge-invariant cubic
vertices  $\sim\varphi\phi\phi$ are forbidden in the theory of massless spin 3
and scalar fields, nevertheless,
local quartic gauge-invariant vertices $\varphi\phi\phi\phi$ can be constructed.
Here, we meet the similar situation, namely, local gauge-invariant quartic vertices
$\sim\varphi A\phi\phi$ can be constructed.
Repeating the main arguments that led to the construction of the generating function
(\ref{c4}) of the anticanonical transformation, the most general form of
the generating function $h^{\mu\nu\lambda}=h^{\mu\nu\lambda}(A,\phi)$
with two derivatives responsible for the generation of quartic  vertices reads
\beq
\nonumber
&&h^{\mu\nu\lambda}=a_1\frac{1}{\Box}
\big[c_1\pa^{(\mu}\pa^{\nu}A^{\lambda)}\;\phi^2+
c_2\pa^{(\mu}A^{\nu}\;\pa^{\lambda)}\phi\;\phi+c_3A^{(\mu}\;\pa^{\nu}\pa^{\lambda)}\phi\;\phi
+c_4A^{(\mu}\;\pa^{\nu}\phi\;\pa^{\lambda)}\phi+\\
\nonumber
&&\qquad\qquad\quad+
c_5\eta^{(\mu\nu}\Box A^{\lambda)}\;\phi^2
+c_6\eta^{(\mu\nu}\pa_{\sigma}A^{\lambda)}\;\pa^{\sigma}\phi\;\phi+
c_7\Box\phi\;\eta^{(\mu\nu}A^{\lambda)}\;\phi+
c_8\eta^{(\mu\nu}A^{\lambda)}\;\pa_{\sigma}\phi\;\pa^{\sigma}\phi+\\
\nonumber
&&\qquad\qquad\quad+c_9\eta^{(\mu\nu}\pa_{\lambda)}\pa_{\sigma}A^{\sigma}\;\phi^2+
c_{10}\eta^{(\mu\nu}\pa^{\lambda)}\phi\;\pa_{\sigma}A^{\sigma}\;\phi+
c_{11}\eta^{(\mu\nu}\pa^{\lambda)}A^{\sigma}\;\pa_{\sigma}\phi\;\phi+\\
&&\qquad\qquad\quad+c_{12}\eta^{(\mu\nu}\pa^{\lambda)}\pa_{\sigma}\phi\;A^{\sigma}\;\phi+
c_{13}\eta^{(\mu\nu}\pa^{\lambda)}\phi\;A^{\sigma}\;\pa_{\sigma}\phi
\big],
\label{c17}
\eeq
where $a_1$ is a coupling constant with ${\rm dim}(a_1)=-(d-2)$ and $c_i,\;\; i=1,2,...,13$
are arbitrary dimensionless constants. For the local addition,
$S_{2\;loc}=S_{2\;loc}[\varphi,A,\phi]$, to the initial action (\ref{c1}),
we obtain
\beq
\nonumber
&&S_{2\;loc}=2a_1\!\!\int \!dx \varphi_{\mu\nu\lambda}
\big[c_1\pa^{(\mu}\pa^{\nu}A^{\lambda)}\;\phi^2+
c_2\pa^{(\mu}A^{\nu}\;\pa^{\lambda)}\phi\;\phi+
c_3A^{(\mu}\;\pa^{\nu}\pa^{\lambda)}\phi\;\phi
+c_4A^{(\mu}\;\pa^{\nu}\phi\;\pa^{\lambda)}\phi-\\
\nonumber
&&\qquad\qquad\quad-(c_1+c_5(d+1))\eta^{(\mu\nu}\Box A^{\lambda)}\;\phi^2
-(c_2+c_6(d+1))\eta^{(\mu\nu}\pa_{\sigma}A^{\lambda)}\;\pa^{\sigma}\phi\;\phi-\\
\nonumber
&&\qquad\qquad\quad-
(c_3+c_7(d+1))\Box\phi\;\eta^{(\mu\nu}A^{\lambda)}\;\phi-
(c_4+c_8(d+1))\eta^{(\mu\nu}A^{\lambda)}\;\pa_{\sigma}\phi\;\pa^{\sigma}\phi-\\
\nonumber
&&\qquad\qquad\quad-(2c_1+c_9(d+1))\eta^{(\mu\nu}\pa^{\lambda)}\pa_{\sigma}A^{\sigma}\;\phi^2-
(c_2+c_{10}(d+1))\eta^{(\mu\nu}\pa^{\lambda)}\phi\;\pa_{\sigma}A^{\sigma}\;\phi-\\
\nonumber
&&\qquad\qquad\quad-
(c_2+c_{11}(d+1))\eta^{(\mu\nu}\pa^{\lambda)}A^{\sigma}\;\pa_{\sigma}\phi\;\phi-
(2c_3+c_{12}(d+1))\eta^{(\mu\nu}\pa^{\lambda)}
\pa_{\sigma}\phi\;A^{\sigma}\;\phi-\\
&&\qquad\qquad\quad-
(2c_4+c_{13}(d+1))\eta^{(\mu\nu}\pa^{\lambda)}\phi\;A^{\sigma}\;\pa_{\sigma}\phi\big].
\eeq
Notice that as in previous case, the gauge generators do not transform under anticanonical
transformations generated by functions (\ref{c17}).

Using integration by parts and performing simple algebraic calculations,
we obtain the variation $S_{2\;loc}$ under gauge
transformations (\ref{c2}) in the form
\beq
\nonumber
&&\delta S_{2\;loc}=-6a_1\!\!\int \!dx\xi_{\nu\lambda}
\Big[-2c_5(d+1)\Box\pa^{\nu}A^{\lambda}\;\phi^2+
(4c_1-c_2-2c_6(d+1))\pa^{\sigma}\pa^{\nu}A^{\lambda}\;\pa_{\sigma}\phi\;\phi-\\
\nonumber
&&\qquad\qquad-(3c_1+2c_9(d+1))\pa_{\sigma}\pa^{\nu}\pa^{\lambda}A^{\sigma}\;\phi^2+
2(c_1-c_2-c_{11}(d+1))\pa^{\nu}\pa^{\lambda}A^{\sigma}\;\pa_{\sigma}\phi\;\phi+\\
\nonumber
&&\qquad\qquad+(c_2-4c_1-4c_5(d+1))\Box A^{\nu}\;\pa^{\lambda}\phi\;\phi-
(c_2-2c_3+2c_6(d+1))\pa^{\sigma}A^{\nu}\;\pa_{\sigma}\pa^{\lambda}\phi\;\phi-\\
\nonumber
&&\qquad\qquad-(c_2-2c_4+2c_6(d+1))\pa^{\sigma}A^{\nu}\;\pa^{\lambda}\phi\;\pa_{\sigma}\phi
-\\
\nonumber
&&\qquad\qquad-
(8c_1+c_2+4c_9(d+1)+2c_{10}(d+1))\pa^{\lambda}\pa_{\sigma}A^{\sigma}\;\pa^{\nu}\phi\;\phi-\\
\nonumber
&&\qquad\qquad-(c_2+4c_3+2c_{11}(d+1)+2c_{12}(d+1))
\pa^{\lambda}A^{\sigma}\;\pa_{\sigma}\pa^{\nu}\phi\;\phi-\\
\nonumber
&&\qquad\qquad-(c_2+4c_4+2c_{11}(d+1)+2c_{13}(d+1))
\pa^{\lambda}A^{\sigma}\;\pa^{\nu}\phi\;\pa_{\sigma}\phi+\\
\nonumber
&&\qquad\qquad+
(c_2-2c_3-2c_7(d+1))\pa^{\nu}A^{\lambda}\;\Box\phi\;\phi+
(c_2-2c_4-2c_8(d+1))\pa^{\nu}A^{\lambda}\;\pa^{\sigma}\phi\;\pa_{\sigma}\phi-\\
\nonumber
&&\qquad\qquad-2c_7(d+1)A^{\lambda}\;\Box\pa^{\nu}\phi\;\phi+
2(c_3-c_4-2c_8(d+1))A^{\lambda}\;\pa_{\sigma}\pa^{\nu}\phi\;\pa^{\sigma}\phi-\\
\nonumber
&&\qquad\qquad-(2c_2-c_3+2c_{10}(d+1))
\pa_{\sigma}A^{\sigma}\;\pa^{\nu}\pa^{\lambda}\phi\;\phi-
(3c_3+2c_{12}(d+1))
A^{\sigma}\;\pa_{\sigma}\pa^{\nu}\pa^{\lambda}\phi\;\phi+\\
\nonumber
&&\qquad\qquad+
(c_3-4c_4-2c_{13}(d+1))A^{\sigma}\;\pa^{\nu}\pa^{\lambda}\phi\;\pa_{\sigma}\phi-
2(c_3-c_4+c_7(d+1))A^{\lambda}\;\pa^{\nu}\phi\;\Box\phi-\\
\nonumber
&&\qquad\qquad-2(2c_3+c_4+c_{12}(d+1)+c_{13}(d+1))
A^{\sigma}\pa_{\sigma}\pa^{\nu}\phi\;\pa^{\lambda}\phi-\\
&&\qquad\qquad-(2c_2-c_4+2c_{10}(d+1))
\pa_{\sigma}A^{\sigma}\pa^{\nu}\phi\;\pa^{\lambda}\phi
\Big],
\label{c21}
\eeq
where the relation $\eta_{\mu\nu}\xi^{\mu\nu}=0$ was systematically used.

Note that the system  of algebraic equations
\beq
\nonumber
&&4c_1-c_2-2c_6(d+1)=0,\quad
3c_1+2c_9(d+1)=0,\quad
c_5=0,\quad
c_1-c_2-c_{11}(d+1)=0,\\
\nonumber
&&c_2-4c_1-4c_5(d+1)=0,\quad
c_2-2c_3+2c_6(d+1)=0,\quad
c_2-2c_4+2c_6(d+1)=0,\\
\nonumber
&&8c_1+c_2+4c_9(d+1)+2c_{10}(d+1)=0,\quad
c_2+4c_3+2c_{11}(d+1)+2c_{12}(d+1)=0,\\
\nonumber
&&c_2+4c_4+2c_{11}(d+1)+2c_{13}(d+1)=0,\quad
c_2-2c_3-2c_7(d+1)=0,\quad c_7=0,\\
\nonumber
&&c_2-2c_4-2c_8(d+1)=0,\quad
2c_2-c_3+2c_{10}(d+1)=0,\quad
3c_3+2c_{12}(d+1)=0,\\
\nonumber
&&c_3-4c_4-2c_{13}(d+1)=0,\quad
c_3-c_4+c_7(d+1)=0,\quad
2c_2-c_4+2c_{10}(d+1)=0,\\
&&2c_3+c_4+c_{12}(d+1)+c_{13}(d+1)=0,\quad
c_3-c_4-2c_8(d+1)=0,
\label{c22}
\eeq
has non-trivial solution
\beq
\nonumber
&&c_2=4c_1,\quad c_3=c_4=2c_1,\quad c_5=c_6=c_7=c_8=0,\\
&&c_9=-\frac{3}{2(d+1)}c_1, \quad c_{10}=c_{11}=c_{12}=c_{13}=-\frac{3}{d+1}c_1 .
\label{c22}
\eeq
Therefore, the functional
\beq
\nonumber
&&S_{2\;loc}=2a_1\!\int \!dx \varphi_{\mu\nu\lambda}
\big[\pa^{(\mu}\pa^{\nu}A^{\lambda)}\;\phi^2+
4\pa^{(\mu}A^{\nu}\;\pa^{\lambda)}\phi\;\phi+
2A^{(\mu}\;\pa^{\nu}\pa^{\lambda)}\phi\;\phi
+2A^{(\mu}\;\pa^{\nu}\phi\;\pa^{\lambda)}\phi-\\
\nonumber
&&\qquad\qquad\quad-\eta^{(\mu\nu}\Box A^{\lambda)}\;\phi^2
-4\eta^{(\mu\nu}\pa_{\sigma}A^{\lambda)}\;\pa^{\sigma}\phi\;\phi-
2\Box\phi\;\eta^{(\mu\nu}A^{\lambda)}\;\phi-
2\eta^{(\mu\nu}A^{\lambda)}\;\pa_{\sigma}\phi\;\pa^{\sigma}\phi-\\
\nonumber
&&\qquad\qquad\quad-\frac{1}{2}\eta^{(\mu\nu}\pa_{\lambda)}\pa_{\sigma}A^{\sigma}\;\phi^2-
\eta^{(\mu\nu}\pa^{\lambda)}\phi\;\pa_{\sigma}A^{\sigma}\;\phi-
\eta^{(\mu\nu}\pa^{\lambda)}A^{\sigma}\;\pa_{\sigma}\phi\;\phi-\\
&&\qquad\qquad\quad-
\eta^{(\mu\nu}\pa^{\lambda)}
\pa_{\sigma}\phi\;A^{\sigma}\;\phi-
\eta^{(\mu\nu}\pa^{\lambda)}\phi\;A^{\sigma}\;\pa_{\sigma}\phi\big].
\eeq
is gauge invariant,
\beq
\delta S_{2\;loc}=0,
\eeq
and presents the quartic vertices.

The local action
\beq
\label{newmod}
S_{loc}[\varphi,A,\phi]=S_0[\varphi,A,\phi]+S_{2\;loc}[\varphi,A,\phi]
\eeq
 describes the model of interacting
$\varphi^{\mu\nu\lambda}$, $A^{\mu}$ and $\phi$ fields
which is invariant under gauge transformations
(\ref{c2})
and belongs to the class of first-stage reducible theories. The action (\ref{newmod})
is described in the explicit form by the functional of finite order in fields
(the fourth in the case under consideration). In turn, this action can be considered as
a simple but not trivial model in the theory of interacting higher spin fields.
The gauge invariance of this action
is checked with the help of
 the explicit, closed and finite (in fields) relation.
 It seems the obtained result cannot be reproduced within the Noether's procedure
 being the main method in studies of interacting higher spin fields
 (see, for example, \cite{Zinov,KMP}).

\section{Gauge-invariant model with mixed vertices}
Results of \cite{L2} and obtained above allow us to formulate the local action
\beq
S[\varphi,A,\phi]=S_0[\varphi,A,\phi]+S_{int}[\varphi,A,\phi]
\eeq
where $S_0[\varphi,A,\phi]$ is defined in (\ref{c1}) and
\beq
\nonumber
&&S_{int}[\varphi,A,\phi]=a_1\!\!\int \!dx\varphi_{\mu\nu\lambda}
\big[\pa^{(\mu}\pa^{\nu}A^{\lambda)}\;\phi^2+
4\pa^{(\mu}A^{\nu}\;\pa^{\lambda)}\phi\;\phi+
2A^{(\mu}\;\pa^{\nu}\pa^{\lambda)}\phi\;\phi
+2A^{(\mu}\;\pa^{\nu}\phi\;\pa^{\lambda)}\phi-\\
\nonumber
&&\qquad\qquad\quad-\eta^{(\mu\nu}\Box A^{\lambda)}\;\phi^2
-4\eta^{(\mu\nu}\pa_{\sigma}A^{\lambda)}\;\pa^{\sigma}\phi\;\phi-
2\Box\phi\;\eta^{(\mu\nu}A^{\lambda)}\;\phi-
2\eta^{(\mu\nu}A^{\lambda)}\;\pa_{\sigma}\phi\;\pa^{\sigma}\phi-\\
\nonumber
&&\qquad\qquad\quad-\frac{1}{2}\eta^{(\mu\nu}\pa_{\lambda)}\pa_{\sigma}A^{\sigma}\;\phi^2-
\eta^{(\mu\nu}\pa^{\lambda)}\phi\;\pa_{\sigma}A^{\sigma}\;\phi-
\eta^{(\mu\nu}\pa^{\lambda)}A^{\sigma}\;\pa_{\sigma}\phi\;\phi-\\
\nonumber
&&\qquad\qquad\quad-
\eta^{(\mu\nu}\pa^{\lambda)}
\pa_{\sigma}\phi\;A^{\sigma}\;\phi-
\eta^{(\mu\nu}\pa^{\lambda)}\phi\;A^{\sigma}\;\pa_{\sigma}\phi\big]+\\
\nonumber
&&\qquad\qquad\quad+b_1\!\int \!dx\varphi_{\mu\nu\lambda}
\big[\pa^{\mu}\pa^{\nu}\pa^{\lambda}\phi \;\phi^2+
2\pa^{(\mu}\pa^{\nu}\phi\;\pa^{\lambda)}\phi\;\phi+
2\pa^{\mu}\phi\;\pa^{\nu}\phi\;\pa^{\lambda}\phi-
\frac{1}{2}\eta^{(\mu\nu}\pa^{\lambda)}\Box\phi\;\phi^2-\\
\label{c23}
&&\qquad\qquad\qquad
-2\eta^{(\mu\nu}\pa^{\lambda)}\pa_{\sigma}\phi\;\pa^{\sigma}\phi\;\phi-
\eta^{(\mu\nu}\pa^{\lambda)}\phi\;
\pa_{\sigma}\phi\;\pa^{\sigma}\phi-
\Box\phi\;\eta^{(\mu\nu}\pa^{\lambda)}\phi\;\phi\big].
\eeq
Here, $a_1$, $b_1$ are coupling constants with
the dimensions ${\rm dim}(a_1)=-(d-2)$, ${\rm dim}(b_1)=-(d-1)$. The action is
invariant under the gauge
transformations (\ref{c2}),
\beq
\delta S[\varphi,A,\phi]=0 .
\eeq
The gauge invariance holds in the limit $m=0$ for scalar field, but it does not maintain for
massless vector field when additional gauge invariance of the initial action appears,
$\delta A_{\mu}=\pa_{\mu}\xi$. In this case, the deformed gauge transformations in the sector
of spin 3 fields become non-local,
\beq
\label{dgt}
\widetilde{\delta}\varphi_{\mu\nu\lambda}=\pa_{(\mu}\xi_{\nu\lambda)}-
h_{\mu\nu\lambda}(A,\phi)\overleftarrow{\pa}_{\!A_{\sigma}}\pa_{\sigma}\xi.
\eeq
In turn, it causes the appearance of additional local part in variation of deformed action
which compensates exactly the variation of $S_{int}[\varphi,A,\phi]$
under gauge transformations of vector field $A^{\mu}$.
Note that the deformed action $\widetilde{S}[\varphi,A,\phi]=S_0[\varphi+h,A,\phi]$
is described by a functional
that has an explicit form and contains vertices up to the eighth order.
 This functional is invariant
under non-local gauge transformations (\ref{dgt}) accompanied
by relations $\widetilde{\delta}A_{\mu}=\pa_{\mu}\xi,$\;$\widetilde{\delta}\phi=0,$\;
$\eta_{\mu\nu}\xi^{\mu\nu}=0$.
These transformations
have an explicit and finite form.

\section{Conclusion}
In the present paper, we have continued our study \cite{L2} in construction
of new gauge-invariant models using recently proposed approach to the deformation procedure
considering as a tool to describe interactions among different fields
\cite{BL-1,BL-2,L-2022}. Our starting point was a free gauge-invariant model
containing  massless spin 3 field,
massive vector and scalar fields. Interaction among the fields was introduced with the help
of a special anticanonical transformation in the BV formalism acting non-trivially in the
sector of spin 3 fields. Moreover, the generating function of the anticanonical
 transformation was chosen to be independent of the spin 3 field. In turn, this led to the
 important consequence that the deformation procedure did not affect
 the original gauge algebra. Further specification of the generating function was dictated
 by our desire to study local cubic and quartic
 vertices containing spin 3 field,  vector field and  one or two scalar fields, respectively.
 Then, it was proved that
 the cubic vertices which must be invariant under
 original gauge transformations due to locality reasons are forbidden while local
 gauge-invariant quartic vertices have been constructed.
The situation is similar
 with the case of local cubic and quartic vertices describing the interactions of the spin 3
 field with the scalar fields studied in \cite{L2}.

Taking into account the results of paper \cite{L2} and those obtained above in Sec. 3, a new
gauge-invariant model with mixed quartic vertices of fields with spins 3, 1 and 0
has been proposed. The action of this model has been described in a closed and explicit form
by a local functional of fourth order in fields. The construction of the model
was completely based on a new approach
to solving the deformation problem using anticanonical transformations in the BV formalism
\cite{BL-1,BL-2,L-2022}. It was noted that this model cannot be reproduced
 within the framework of the standard Noether procedure,
which is the main method for studying interactions in the theory of higher spin fields.

\section*{Acknowledgments}
\noindent
The author thanks I.L. Buchbinder for useful discussions.
The work is supported by the Ministry of
Education of the Russian Federation, project FEWF-2020-0003.

\end{document}